\documentclass [preprint,12pt]{elsarticle}

\usepackage{graphicx}
\usepackage{epstopdf}

\usepackage{amssymb}

\journal{Astroparticle Physics}

\begin{document}

\begin{frontmatter}



\title{Gamma-ray signatures of cosmic ray acceleration, propagation, and confinement in the era of CTA}


\author[Montpellier]{F. Acero}
\author[Tokyo]{A. Bamba}
\author[APC,MPIK,Bochum]{S. Casanova\corref{cocorr}}
\author[Barcelona]{E. de Cea}
\author[MPIK]{E. de O\~na Wilhelmi}
\author[APC]{S. Gabici\corref{corr}}
\author[Montpellier]{Y. Gallant}
\author[Barcelona]{D. Hadasch}
\author[Montpellier]{A. Marcowith}
\author[Barcelona]{G. Pedaletti}
\author[Innsbruck]{O. Reimer\corref{cocorr}}
\author[Montpellier]{M. Renaud}
\author[Barcelona,Barcelona2]{D. F. Torres}
\author[MPIK]{F. Volpe}
\author{for the CTA collaboration.}

\cortext[corr]{Principal corresponding author, stefano.gabici@apc.univ-paris7.fr, tel. +33157277024, fax. +33157276071}
\cortext[cocorr]{Corresponding author}

\address[APC]{Astroparticule et Cosmologie (APC), CNRS, Universit\'e Paris 7 Denis Diderot, 10 rue Alice Domon et Leonie Duquet, F-75205 Paris Cedex 13, France}
\address[MPIK]{Max-Planck-Institut f\"ur Kernphysik, Saupfercheckweg 1, 69117 Heidelberg, Germany}
\address[Bochum]{Unit for Space Physics, North-West University, Potchefstroom 2520, South Africa}
\address[Innsbruck]{Institut f\"ur Astro- und Teilchenphysik, Leopold-Franzens-Universit\"at Innsbruck A-6020 Innsbruck, Austria}
\address[Montpellier]{LUPM - UMR5299, Universit\'e de Montpellier II, Place Eug\`ene Bataillon - CC 72, 34095 Montpellier C\'edex 05 FRANCE}
\address[Barcelona]{Institut de Ci\`encies de l'Espai (IEEC-CSIC), Campus UAB,  Torre C5, 2a planta, 08193 Barcelona, Spain}
\address[Barcelona2]{Instituci\'o Catalana de Recerca i Estudis Avan\c{c}ats (ICREA)}
\address[Tokyo]{College of Science and Engineering, Aoyama Gakuin University,
5-10-1 Fuchinobe, Chuo-ku, Sagamihara,
Kanagawa, 252-5258, Japan}

\begin{abstract}
Galactic cosmic rays are commonly believed to be accelerated at supernova remnants via diffusive shock acceleration. Despite the popularity of this idea, a conclusive proof for its validity is still missing. Gamma-ray astronomy provides us with a powerful tool to tackle this problem, because gamma rays are produced during cosmic ray interactions with the ambient gas. The detection of gamma rays from several supernova remnants is encouraging, but still does not constitute a proof of the scenario, the main problem being the difficulty in disentangling the hadronic and leptonic contributions to the emission.
Once released by their sources, cosmic rays diffuse in the interstellar medium, and finally escape from the Galaxy. The diffuse gamma-ray emission from the Galactic disk, as well as the gamma-ray emission detected from a few galaxies is largely due to the interactions of cosmic rays in the interstellar medium.
On much larger scales, cosmic rays are also expected to permeate the intracluster medium, since they can be confined and accumulated within clusters of galaxies for cosmological times.
Thus, the detection of gamma rays from clusters of galaxies, or even upper limits on their emission, will allow us to constrain the cosmic ray output of the sources they contain, such as normal galaxies, AGNs, and cosmological shocks.
In this paper, we  describe the impact that the Cherenkov Telescope Array, a future ground-based facility for very-high energy gamma-ray astronomy, is expected to have in this field of research.
\end{abstract}


\end{frontmatter}


\section{Introduction}
\label{sec:intro}

According to the most popular scenario, galactic cosmic rays (CRs) are accelerated at supernova remnants (SNRs) via diffusive shock acceleration (for a review, see e.g. \cite{hillas}). The main argument supporting this scenario is the fact that SNRs alone would be able to maintain the CR population at the observed level, if some fraction (some 10\%) of their kinetic energy were somehow converted into CRs. Moreover, diffusive shock acceleration predicts a spectral shape for the accelerated particles that, after taking into account propagation effects in the Galaxy, fits fairly well with the observed CR spectrum, and, under certain assumptions for particle injection, is roughly compatible with the observed chemical composition of CRs (see, e.g., \cite{ptuskinspectrum} and references therein). All these facts are encouraging, but an unambiguous and conclusive proof for (or against) the validity of this scenario is still missing. The main problem in this respect is connected to the isotropy of the arrival directions of CRs in the sky. This is due to their diffusive behavior in the galactic magnetic field, which prevents the identification of CR sources as is done for sources of photons in astronomy.

Very tight connections exist between CR studies and gamma-ray astronomy, due to the fact that CR protons can undergo hadronic interactions with the interstellar medium producing neutral pions that in turn decay into gamma rays (e.g. \cite{stecker,dermer}). This is of particular relevance for the identification of CR sources, since the production of gamma rays is expected, at some level, during CR acceleration. If SNRs indeed are the sources of CRs, they have to convert $\sim 10\%$ of their explosion energy into accelerated particles. Since the explosion energy of a supernova is a remarkably constant quantity close to $10^{51}$~erg, a rough estimate of the expected gamma-ray flux from a given SNR can be obtained if one knows the density of the ambient medium, and the SNR distance. Such estimates fall within the capabilities of currently operating Cherenkov telescopes \cite{dav,naitotakahara}. In agreement with these early estimates, several SNRs have been detected at TeV energies (for recent reviews see e.g. \cite{felixreview,hofmannreview}) and, though their detection fits well with the general scenario described above, the origin of such radiation might still be leptonic, and thus unrelated to the acceleration of hadronic CRs. In particular, for the best studied SNR in TeV gamma rays, RX~J1713.7-3946 \cite{RXJ1713}, evidence has been put forward, both from X-ray \cite{donleptonic} and GeV gamma-ray \cite{fermiRXJ} observations, supporting a leptonic origin of the TeV emission. This does not mean that the SNR RX J1713.7-3946 is not accelerating hadronic CRs at the level required to explain the flux of galactic CRs, but simply that the leptonic contribution to the gamma ray production is dominant over the contribution from hadrons. This may naturally happen if the ambient gas density is low. On the other hand, the gamma-ray emission detected from the historical SNR Tycho, seems to favor an hadronic origin \cite{fermitycho} (but see \cite{dermertycho} for an alternative interpretation). Further multiwavelength studies of SNRs are needed in order to find a conclusive evidence of the fact that SNRs, as a class, accelerate hadronic CRs and are quantitatively able to provide the measured flux of CRs.

At some stage of the SNR evolution, CRs have to be released into the interstellar medium. The details of the process of particle escape from SNRs are still to be fully understood, though it is generally believed that particles with the highest energies escape first, while particles with lower and lower energies are gradually released as the SNR shock speed decreases \cite[e.g.][]{ptuskinescape}.
Once escaped, CRs diffuse away from the acceleration site and, before being dispersed in the galactic CR background, they remain for some time diffusively confined in the vicinity of the accelerator.
The details of CR diffusion in the interstellar medium are also not fully understood (see e.g. \cite{lazarian} and references therein). However, by assuming that CR diffusion proceeds isotropically, and that at a given time CRs have diffused away from the source and fill a region of radius $R$, it is possible to estimate the average energy density of CRs as \cite{atoyan}:
$
w_{CR} \approx 0.55 ~ (W_{CR}/10^{50}{\rm erg})(R/100~{\rm pc})^{-3} {\rm eV/cm^3}
$
, where $W_{CR}$ is the total energy carried by CRs when released by the source.
This means that, in regions up to 100 pc or so away from SNRs (or from any other source injecting $10^{50}$~erg in form of CRs), and at some given time, depending on the details of CR escape and diffusion, the CR intensity might well be above the background intensity of galactic CRs, which is $\approx 1~{\rm eV/cm^3}$. As a consequence, the gamma-ray emission from CR interactions also would be correspondingly enhanced. Thus, searching for excesses of gamma-ray emission from the regions surrounding candidate CR sources might provide an indirect way to identify and locate CR sources. The detection of this radiation is more likely if massive molecular clouds are located within the region filled by CRs, since they would provide a thick target for CR hadronic interactions \cite{atoyan,gabici07}. Such radiation might have been already detected at TeV energies from some regions in the Galaxy, including the massive molecular clouds located within $\approx 100$~pc from the galactic centre (the SNR Sgr A East might be one of the potential sources of CRs in that region)  \cite{HESSridge} and the molecular clouds located in the vicinity of the SNR W28 \cite{W28,lichen,ohira,gabiciW28}. However, further detections are needed in order to test the reliability of such interpretation.

CRs escaping from SNRs and from any other CR source in the Galaxy eventually mix with the CR background and sustain it against CR escape from the Galaxy. Except for localized (both in time and space) excesses around CR sources, the CR intensity is expected to be, both spatially and temporally, quite homogeneous throughout the Galaxy. The interactions between CRs and the interstellar gas makes the galactic disk a prominent source of diffuse gamma rays at energies above $\approx 100$~MeV \cite{FERMIdiffuse}. No diffuse emission has been firmly detected at TeV energies from the galactic disk, though some evidence for the presence of diffuse emission at $\approx 15$~TeV has been presented by the MILAGRO collaboration \cite{MILAGROdiffuse}. Studying the diffuse emission is of crucial importance in order to probe large scale spatial variations in the distribution of CRs and thus constrain the properties of their propagation in the turbulent galactic magnetic field.

The detection by the Fermi satellite of gamma rays from nearby galaxies \cite{FERMILMC,FERMISMC,FERMIlocal}, as well as the detection in both the GeV and TeV energy range of the starburst galaxies NGC 253 and M82 \cite{FERMIstarbursts,HESSstarburst,VERITASstarburst} also can be interpreted as the result of CR interactions with the interstellar gas.
Similarly to what it is done for our Galaxy, such gamma-ray emission can be used to infer the CR intensities in these objects.
Thus, the study of this emission and the detection of more galaxies in gamma rays will serve to investigate the possible differences in the acceleration of CRs in galaxies different from our own, and in their transport and confinement properties.
The lessons of these systems will also indirectly contribute to solving the problem of the origin of galactic CRs.

Moving to even larger scales, one may wonder where CRs end up after escaping the galaxy within which they have been accelerated. If the parent galaxy is a member of a group or a cluster of galaxies, CRs will remain confined for cosmological times in the magnetized ($\approx 0.1 ... 1 ~ \mu{\rm G}$) intracluster medium.
This is because of the very large, Mpc-scale size of these objects that makes the confinement time of CRs larger than the Hubble time \cite{vab,bbp}  (see \cite{ensslin} for a recent discussion of CR confinement in clusters of galaxies).
As a consequence, all the CRs injected by all the sources within a cluster of galaxies accumulate in the intracluster medium, and for this reason clusters of galaxies often have been considered as potential gamma-ray sources.
The reason is that, despite the very low ambient density ($\approx 10^{-4} ~ {\rm cm}^{-3}$), one may still expect a copious production of gamma rays from proton-proton interactions due to the huge amount of CRs that possibly can be stored within clusters (for a review see \cite{reviewclusters} and references therein).
To date no cluster of galaxies has been detected in GeV nor TeV gamma rays \cite[e.g.][]{FERMIclusters,HESScoma,MAGICperseus} and the upper limits on the gamma-ray fluxes can be used to constrain the total power of all the CR sources hosted by clusters. The list of such sources includes, besides normal and starburst galaxies, active galactic nuclei and cosmological shock waves.

The Cherenkov Telescope Array \cite{CTA}, currently in the design phase, is planned to be the most sensitive array operating in the TeV energy domain.
The sensitivity at energies $\gtrsim 100 ~ {\rm GeV}$ is expected to be improved by a factor of 5-10 with respect to currently operating instruments, and the accessible energy range to be extended down to a few tens of GeV and up to $\approx$~100 TeV.
Such an instrument will considerably increase the number of detected sources (of particular relevance here are SNRs, localized molecular clouds, and nearby galaxies), facilitate population studies, and potentially detect new classes of sources, such as clusters of galaxies at gamma-ray energies. 
It is important to remind that both SNRs and MCs are extended object. Thus the large field of view of CTA ($\approx$5-10$^{\circ}$), coupled with an improved angular resolution ($\approx$0.05$^{\circ}$ above 1 TeV) will allow us to perform imaging studies with unprecedented accuracy. This is of particular relevance, for example, in order to identify SNRs based on their shell-like morphology, or to perform correlation studies between TeV maps and maps obtained in other energy bands, most notably in the X-ray domain.
The large field of view will also facilitate the study of SNR/MC associations, namely MCs which are illuminated by CRs coming from nearby SNRs.
A determination, with high sensitivity, of spatially resolved gamma-ray sources related to
the same accelerator would lead to the experimental determination of the local diffusion
coefficient and/or the local injection spectrum of cosmic rays.
To conclude, the large effective area coupled with an excellent energy resolution ($\approx$10\% above 1 TeV) will allow us to determine with great accuracy sources' spectral parameters. This will in turn constrain the spectrum of the CRs which are producing the observed radiation. These high quality spectral data, combined with the data available in the GeV energy range (i.e. from the Fermi and Agile satellites), will allow to perform a detailed spectral modeling and possibly resolve the ambiguity between hadronic and leptonic emission from SNRs.
All these studies will constitute a large advance for understanding the acceleration, transport and confinement of CRs in our Galaxy and more generally in galactic systems. The aim of this paper is to discuss the role that CTA will play in driving this field of research by presenting detailed simulations of CTA observations for a number of selected science cases.

This is the first paper fully devoted to the description of the impact that the Cherenkov Telescope Array will have on future CR studies. Previous general discussions on the role of CTA that also mentioned CR studies can be found in \cite{CTA,matthieu}.

In the following discussion, several of the configurations (i.e. number of small/middle/large telescopes, their spatial distribution, etc) proposed for CTA will be investigated. Different configurations will result in different instrumental characteristics such as effective area, angular resolution, or sensitivity. Here, we use configurations B, D, and I (or E) as representative of the configurations which are characterized by a sensitivity optimized in the low, high, and over a broad energy range, respectively. A description of the different configurations can be found in a companion paper in this issue \cite{CTAconfigurations}. Here we briefly summarize the main differences between the proposed configurations for CTA. For all configurations, large size telescopes (LSTs, dish $\sim$ 23 m) will be placed at the center of
the array. Thanks to their large mirror area, dim flashes of Cherenkov light
from the low energy events ($\sim$ 50 GeV) are expected to be reconstructed. Tens
of medium size telescopes (dish $\sim$ 11m) will be placed in a surrounding
ring, covering a large fraction of the light pool and thus enhancing the
reconstruction of medium energy ($\sim$ 1 TeV) events. Finally, an outer ring
will be composed of small size telescopes (SSTs, dish $\sim$ 7 m) enlarging the
effective area of the array for the bright but rare high energy events
(above $\sim$ 50 TeV). While this characteristic stay for most of the
configurations explored, the simulated arrays can be divided in 3 main
classes of constant estimated cost: low energy focus, well balanced, high
energy focus. In the low energy focus (configurations A,B,F,G), the number of LSTs is increased at the expenses of SSTs, depleting sensitivity at energies 10TeV. As the name of the class suggests, the well balanced type array
(configurations E,I,J,K) reaches a compromise in performances in the whole energy range and the high energy focus one (configurations C,D,H) increase the number of SSTs with little or no LSTs.

The paper is structured as follows: galactic sources (namely supernova remnants and molecular clouds) are discussed in Section 2, extragalactic sources (namely starburst galaxies and clusters of galaxies) are discussed in Section 3. Conclusions are in Section 4. The paper is mainly focused on the hadronic component of CRs, which constitute the bulk of the cosmic rafdiation. For a brief and recent review on CR electrons see \cite{CRelectrons}. 

\section{Supernova remnants and molecular clouds}

\subsection{Population studies of supernova remnants with CTA}

Being characterized by an improvement of a factor of 5-10 in sensitivity with respect to currently operating telescopes, CTA is expected to detect a large number of new gamma-ray sources.
To date, only 5 of the SNRs that have been detected in TeV gamma rays exhibit a clear shell-like morphology: RX~J1713.7-3946 \cite{RXJ1713}, Vela Junior \cite{velajunior}, RCW 86 \footnote{Note that in this case a shell-like morphology fit on the source radial profile is only slightly favored statistically over a simple sphere fit.} \cite{RCW86}, SN 1006 \cite{SN1006}, and HESS J1731-347 \cite{J1731}. Other SNRs have been detected, but their morphology remains unresolved (for an updated list see e.g. \cite{damiano}).
However, the total number of detected objects is still too small to allow population studies or a satisfactory statistical analysis of the SNR properties in gamma rays.
The situation is expected to change significantly after CTA will begin to take data, since the number of SNRs detected in TeV gamma rays will undoubtedly increase significantly.

A rough estimate of the total number of SNRs that CTA will detect can be computed as follows: let us assume that the SNRs detected so far in TeV gamma rays are good representatives of the whole class of SNRs. In other words, let us assume that the gamma-ray luminosities of these objects (or their average luminosity) can be considered as typical luminosities for all SNRs. Then, from the expected sensitivity of CTA we can derive the maximum distance at which a generic SNR would be detected. We call this distance the {\it horizon of detectability}. Finally, from the knowledge of the spatial distribution of supernovae in the Galaxy \cite{case}, their explosion rate \cite{SNrate}, and the duration of the TeV emission (believed to last a few thousand years, see e.g. \cite{ptuskinescape}) we can obtain the prediction for the number of objects detectable by CTA.

Normally, TeV-bright SNRs are detected by pointing the telescope in the direction of SNRs known from other wavelengths. However, SNR detection can also happen in reverse order, as in the case of the SNR HESS J1731-347, where the discovery of the radio shell follows the SNR detection in gamma rays \cite{J1731}.
Similar situations are expected to happen frequently once the number of detections has increased.
Of course, if the gamma-ray source clearly exhibits a shell-like structure, it can be confidently identified as an SNR, while for unresolved sources follow-up at other wavelengths is mandatory.
For this reason, it seems appropriate to define a {\it horizon of resolvability}, defined as the maximum distance up to which the shell of an SNR can be spatially resolved and distinguished from a simple gaussian shape \citep{matthieu}.
Then one can proceed as explained above and compute the number of SNRs detectable {\it and} resolvable by CTA.

Note that here we limit ourselves to consider {\it isolated} SNRs only, i.e. SNRs which are not interacting with a molecular clouds. One of the best studied objects belonging to this category is IC 443, which is discussed in Sec.~\ref{sec:spectral}. An extended discussion on SNR/MC associations can be found in Sec.~\ref{sec:SNRMC}.

\begin{figure}
\centering
\includegraphics[width=.8\textwidth]{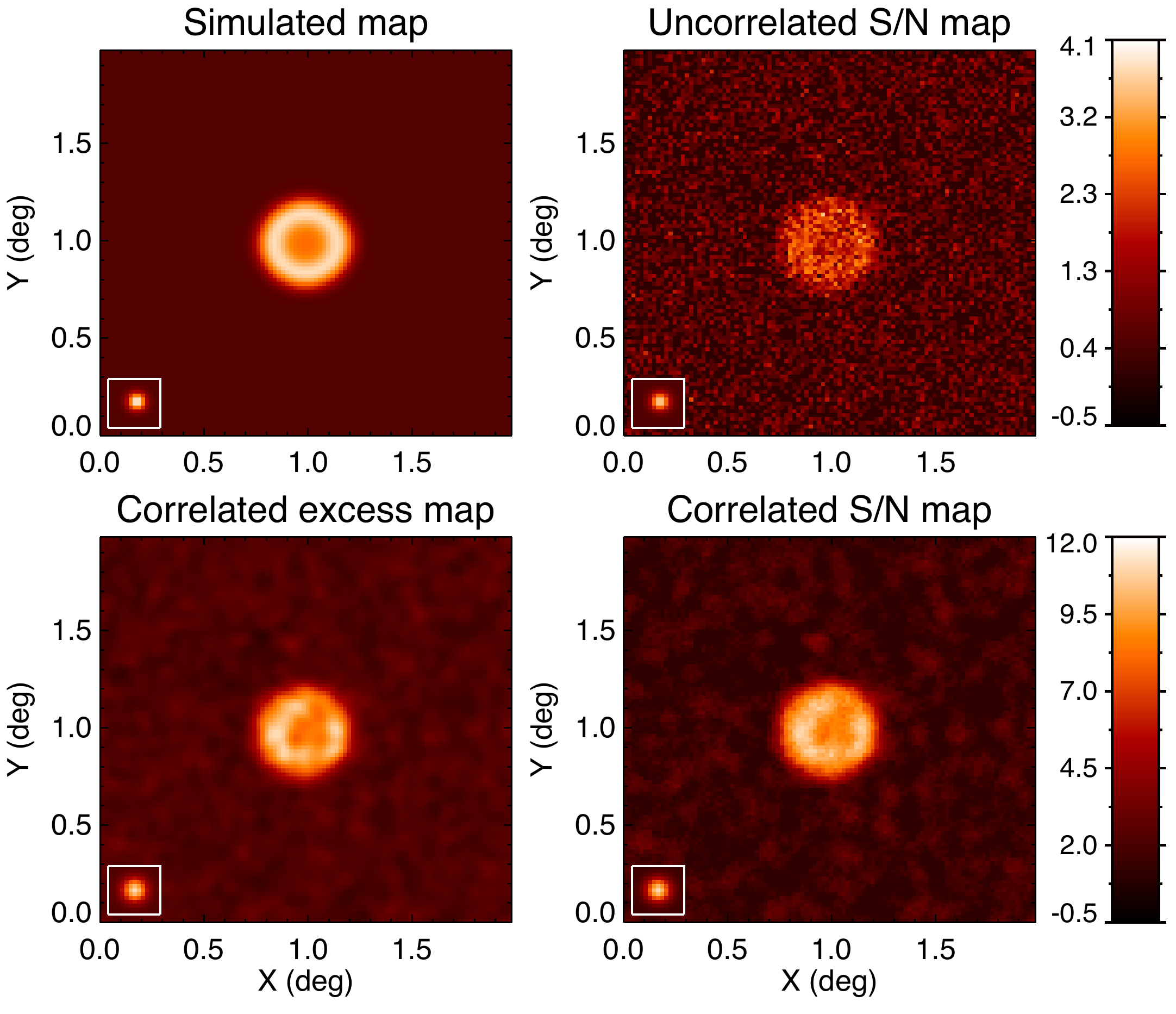}
\caption{Simulated image of a RX J1713-like SNR located at a distance of 3 kpc. Configuration I
has been considered for 20h of exposure time. The PSF (68\% containment radius) is shown in the bottom-left corner of each
panel. The four panels show, from top-left clockwise, the simulated map, the signal-to-noise ratio per pixel, the correlated signal-to-noise map, and the
correlated excess map, respectively. See text and \cite{matthieu} for more details.}
\label{SNRmap}
\end{figure}

\begin{figure}
\centering
\includegraphics[width=.8\textwidth]{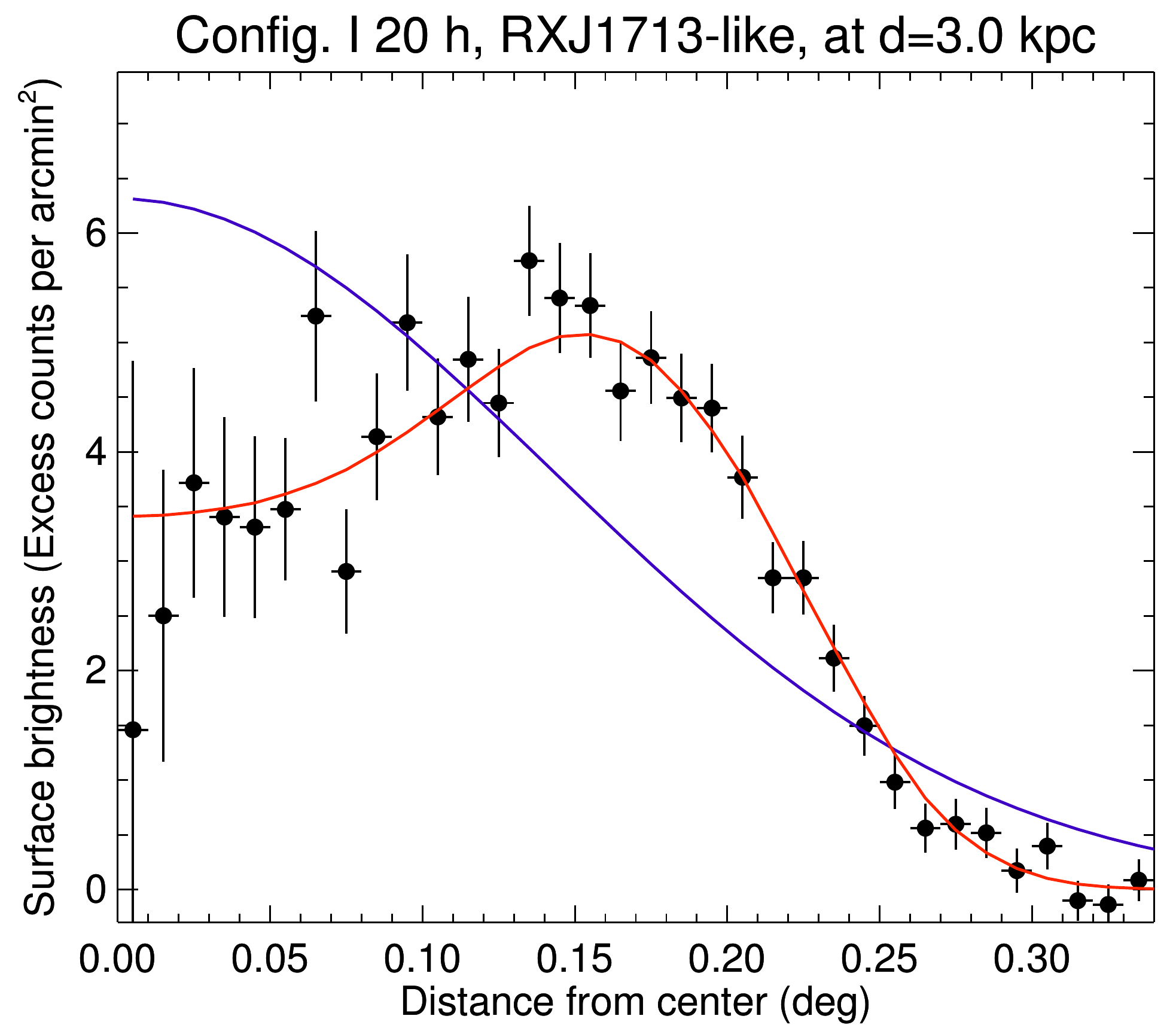}
\caption{Radial profile of the RX J1713-like SNR from Figure~\ref{SNRmap} (data points) fitted with a gaussian profile (blue line) and a shell (red line). The latter fit is clearly favored.}
\label{shellfit}
\end{figure}

Before proceeding to such estimates, it is important to investigate the capability of the angular resolution of CTA in identifying SNRs based on their shell-like morphology.
For this purpose, we show in Figure~\ref{SNRmap} the simulated image of a RX~J1713-like SNR located at a distance of 3 kpc (the actual distance would be $\approx$ 1 -- 1.5 kpc).
Configuration I has been used, with 20h of observations.
The different panels show the simulated map (top-left), the signal-to-noise ratio per pixel (top-right), the
uncorrelated (i.e. per pixel) excess map, correlated with a 2-dimensional gaussian PSF (bottom-left), and finally the signal-to-noise ratio, calculated on a circular region whose root-mean-square radius is the same as for the PSF (bottom-right).
The PSF is shown in the bottom-left corner of each panel and has a size of 0.04 degrees.
The maps' pixel size is 0.01 degrees.
At a distance of 3 kpc the shell structure is clearly still
visible. This is evident from Figure~\ref{shellfit}, where the surface brightness profile for the RX
J1713-like SNR is fitted with a gaussian (blue line) and with a shell (red line).
The shell morphology is clearly favored.
This method has been used to compute the
resolvability horizon of SNRs described above.

\begin{figure}
\centering
\includegraphics[width=\textwidth]{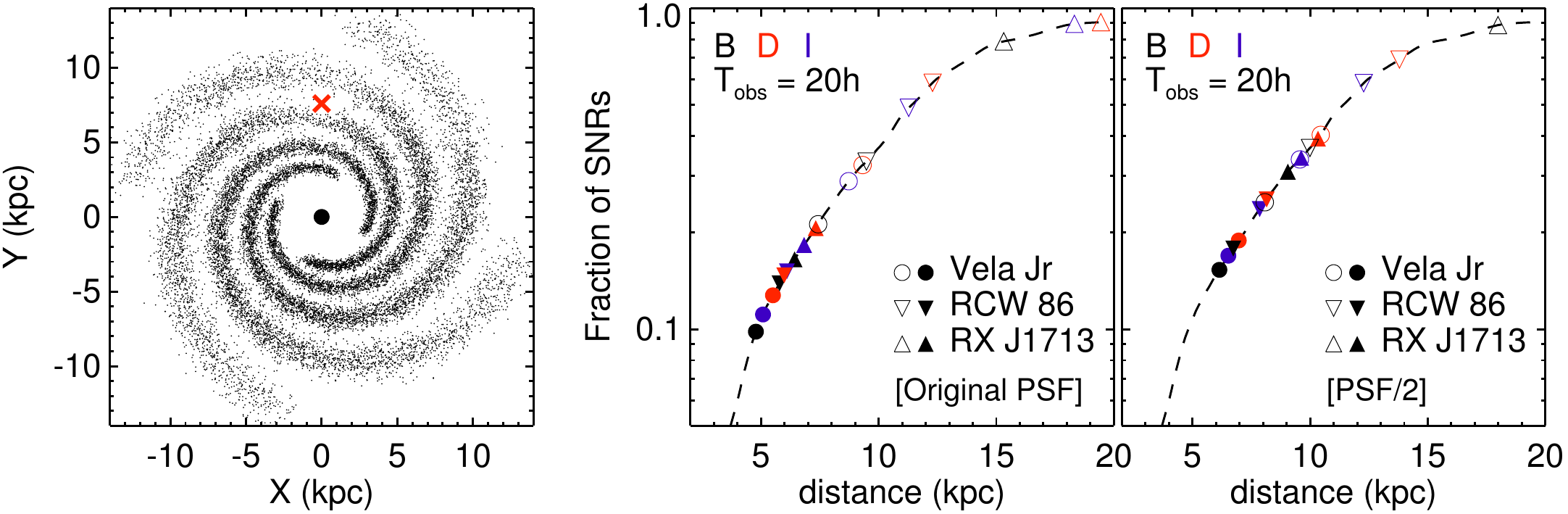}
\caption{Left: simulated distribution of Galactic core-collapse SNRs.  The Sun and the Galactic center are depicted by the red
cross and the black dot, respectively.
Middle: Fraction of SNRs visible with zenith angle $< 45^{\circ}$,
for a ground-based instrument located at the same latitude as the H.E.S.S. experiment, as a
function of distance from Earth (dashed line). Configurations B (in black), D (in red) and I (in blue), for an
observing time of 20 h, and for three known VHE-emitting SNRs: Vela Junior (circles), RCW 86
(downward triangles) and RX J1713.7-3946 (upward triangles). Open symbols refer to the horizon
of detectability, filled symbols to the horizons of resolvability. Right: same as middle, except that the original
CTA PSF from the configuration files has been improved by a factor of 2.}
\label{detectability}
\end{figure}

We can now proceed to compute the expected number of SNRs detectable and resolvable by CTA.
The main results are shown in Figure~\ref{detectability}. 
The simulated spatial distribution of SNRs in the galaxy is shown in the left panel. A logarithmic model of spiral arms of \cite{vallee} has been adopted, with an arm dispersion which is a function of the distance from the Galactic center \cite{drimmel}. The SNR Galactocentric distribution is
provided in \cite{case}.
The Sun and the Galactic center are depicted by the red cross and the black dot, respectively.

The fraction of SNRs located within a given distance from the Sun and visible with zenith angle $< 45 ^{\circ}$, for a ground-based instrument located in the southern hemisphere, at the same latitude as the H.E.S.S. experiment is plotted as a dashed line in
the middle and right panels of Figure~\ref{detectability}.
The \emph{horizon of detectability} and \emph{of resolvability} have been computed for three TeV-bright shell type SNRs: Vela~Jr (circles), RCW 86 (downward triangles), and RX~J1713 (upward triangles). The \emph{horizon of resolvability} is indicated with the filled symbols
and is defined as the distance up to which the shell-like morphology of those objects would
be significantly identified by CTA and favored at 3~$\sigma$ over the uniform sphere model. The \emph{horizon of detectability} is indicated by the open
symbols and indicates the maximum distance up to which the three SNRs would be
detectable by CTA with a peak significance of 5~$\sigma$, regardless of their morphology.
The two horizons have been defined after simulating 100 SNR images per distance bin for an observing time of 20 hours.
Different colors refer to three different configurations of the array: B (black), D (red), and I (blue).

If the three objects considered here indeed are typical representatives of
the whole SNR class, the results from Figure~\ref{detectability} could be considered as a valid estimate
of the actual fraction of galactic SNRs detectable by CTA. If we assume that an SNR is
bright in TeV gamma rays for $\sim$~3000 yr (this is approximatively the age of Vela Jr), and we
recall that $\sim$~2.8 supernovae are expected to explode each century in the Galaxy \cite{SNrate}, we can infer that the number of SNRs currently emitting TeV gamma rays is $\sim$~80.
One can then use the results from Figure~\ref{detectability} (middle and right panel) to infer the number of
SNRs detectable (or resolvable) by CTA.
The difference between the middle and the right panel is in the PSF that has been
assumed in the calculations. In the middle panel the 68\% containment radius of the PSF is
the one estimated for the given configuration of CTA, while in the right panel is half this value. Thus, the difference between the two plots demonstrates how an improvement in the instrument angular resolution would affect the results. While the number of
detectable SNRs slightly increases by a factor of $< 1.2$, the gain factor in the number of resolvable
SNRs ranges from 1.5 to 1.9.
In producing Figure~\ref{detectability} we assumed a zenith angle of 20$^{\circ}$. The PSF is
assumed to be gaussian.
It is clear from Figure~\ref{detectability} that CTA will have the capability to detect SNRs as luminous as RX J1713, Vela Jr, or RCW 86 up to the other side of the Galaxy. So, if the assumptions made here are correct, virtually all the TeV bright SNRs in the Galaxy will be detected.
However, for distances larger than the horizon of resolvability, identifying an object as an SNR might be difficult in the absence of observations at other wavelengths. As a consequence, a significant fraction of the SNRs that will be detected by CTA will be classified at first as unidentified sources.

 As a consistency check, we also computed the horizon of detectability of the SNRs Cas~A and Tycho, which are also detected at both GeV and TeV energies, like the SNRs considered in Fig.~\ref{detectability}, but are significantly younger, with an age of few hundreds years. For CTA in configuration I and for an observation time of 20 hours, the horizon of detectability for these two objects is $\sim$9 kpc and $\sim$5 kpc, respectively. Such numbers are comparable with the ones reported in Fig.~\ref{detectability}. Since Cas~A and Tycho are not resolved in gamma rays, the horizon of resolvability cannot be computed. 

The key observational parameters characterizing the SNRs that have been used to compute the horizons are summarized in Table~\ref{table}.

\begin{table}[b!]
\centering
\begin{tabular}{|c|c|c|c|c|c|c|}
\hline
Name & $t_{age}$ & d & $N_0(10^{-12}/$ & $\Gamma$ & $E_c$ & $\beta$  \\
  & ({\rm kyr}) & ({\rm kpc})& ${\rm cm^2/s/TeV}$)     &          &  ({\rm TeV}) &  \\ 
\hline
Vela Jr & 1.7-4.3 & 0.8 & 35.0 & 1.80 & 3.0 & 0.5  \\
RCW 86 & 1.8 & 2.5 & 3.72 & 2.54 & -- & --  \\
RXJ1713 & 1.6 & 1.3 & 58.2 & 1.62 & 0.78 & 0.38 \\
Cas~A & 0.33 & 3.4 & 4.46 & 2.10 & 0.625 & 0.5   \\
Tycho & 0.44 & 3.0 & .126 & 2.30 & -- & --  \\
\hline
\end{tabular}
\caption{Observational parameters for the SNRs considered in the text. The first three columns refer to SNR name, age, and distance, respectively. The columns 4-7 refer to the best fit spectral parameters to the FERMI and TeV data, assuming that the differential spectrum has the shape $N = N_0 \times (E/{\rm TeV})^{-\Gamma} \exp{-(E/E_c)^{\beta}} {\rm cm^{-2} s^{-1} TeV^{-1}}$.}
\label{table}
\end{table}

\subsection{Spectral studies of supernova remnants with CTA: probing cosmic ray acceleration}
\label{sec:spectral}

Another important issue to be investigated is the capability of CTA to perform spectral studies of SNRs and identify possible spectral features such as breaks or cutoffs.
This is of crucial importance in order to determine with good accuracy the spectral shape of the underlying CR population (be they hadrons or electrons) which are responsible for the gamma-ray emission.
To date, a spectral cutoff in the TeV range has been clearly identified only in the spectrum of the bright SNR RX~J1713.7-3946 \cite{RXJ1713cutoff}, for all the other (fainter) TeV-bright SNRs the statistics at high energies does not permit to draw firm conclusions on the spectral shape.
Determining the cutoff energy in the gamma-ray spectrum is required in order to obtain information about the maximum energy of the particles accelerated at SNR shocks.
Moreover, in the context of diffusive shock acceleration, the shape of the cutoff in the proton or electron spectrum is different if the maximum energy is determined by particle escape, by the age of the accelerator, or by possible particle energy-loss processes. Also, the dependency on energy of the particle diffusion coefficient at the shock shapes the resulting spectrum of CRs, the stronger the dependence, the sharper the cutoff. These differences are also present in the resulting radiation spectrum. Thus, high quality spectra can be used to constrain the properties of the acceleration mechanism operating at SNR shocks, or to distinguish between an hadronic or leptonic origin of the gamma-ray emission (see e.g. \cite{morlinoRXJ1713,vladRXJ1713}).

\begin{figure}
\centering
\includegraphics[width=0.6\textwidth]{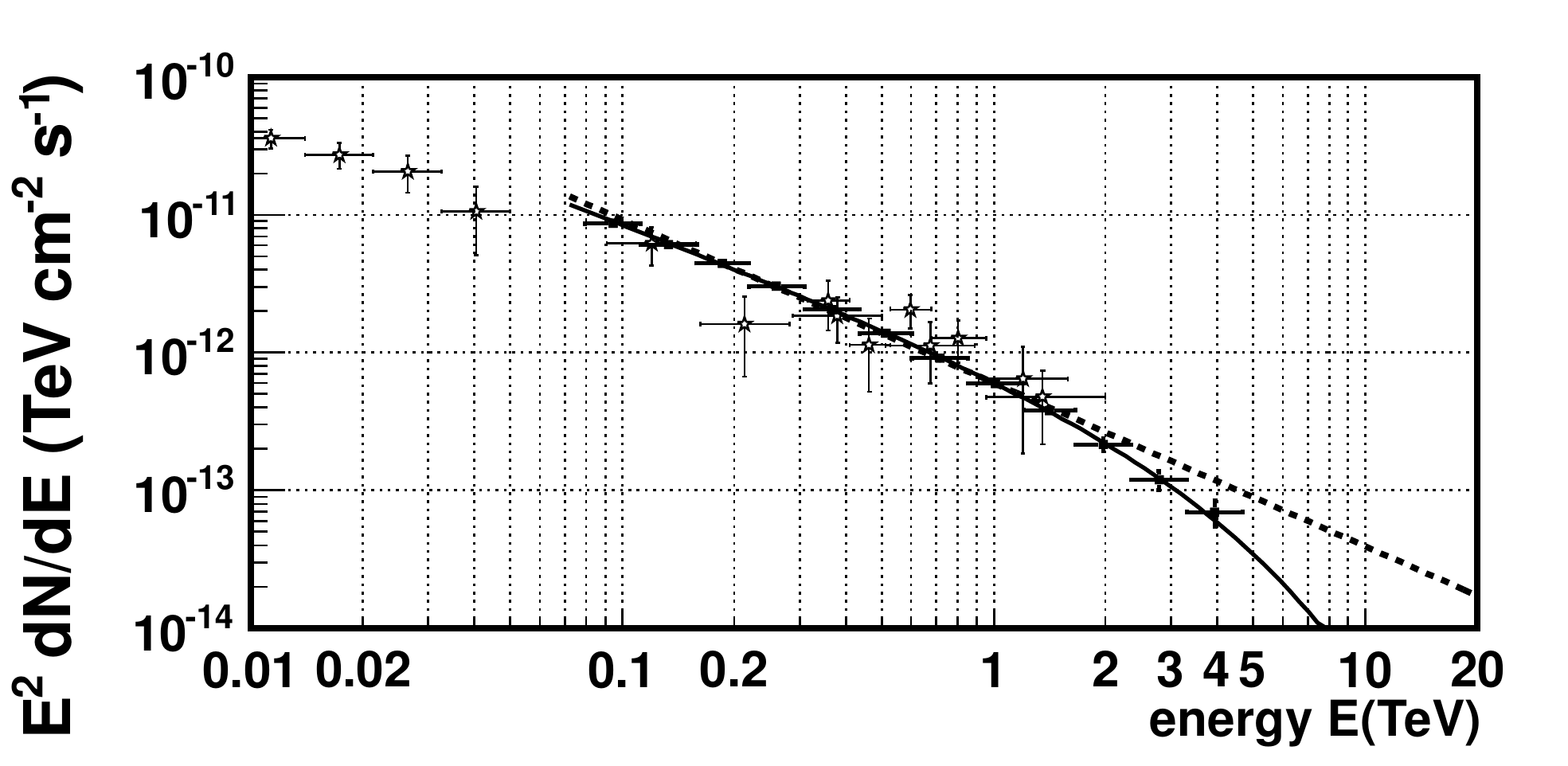}
\includegraphics[width=0.6\textwidth]{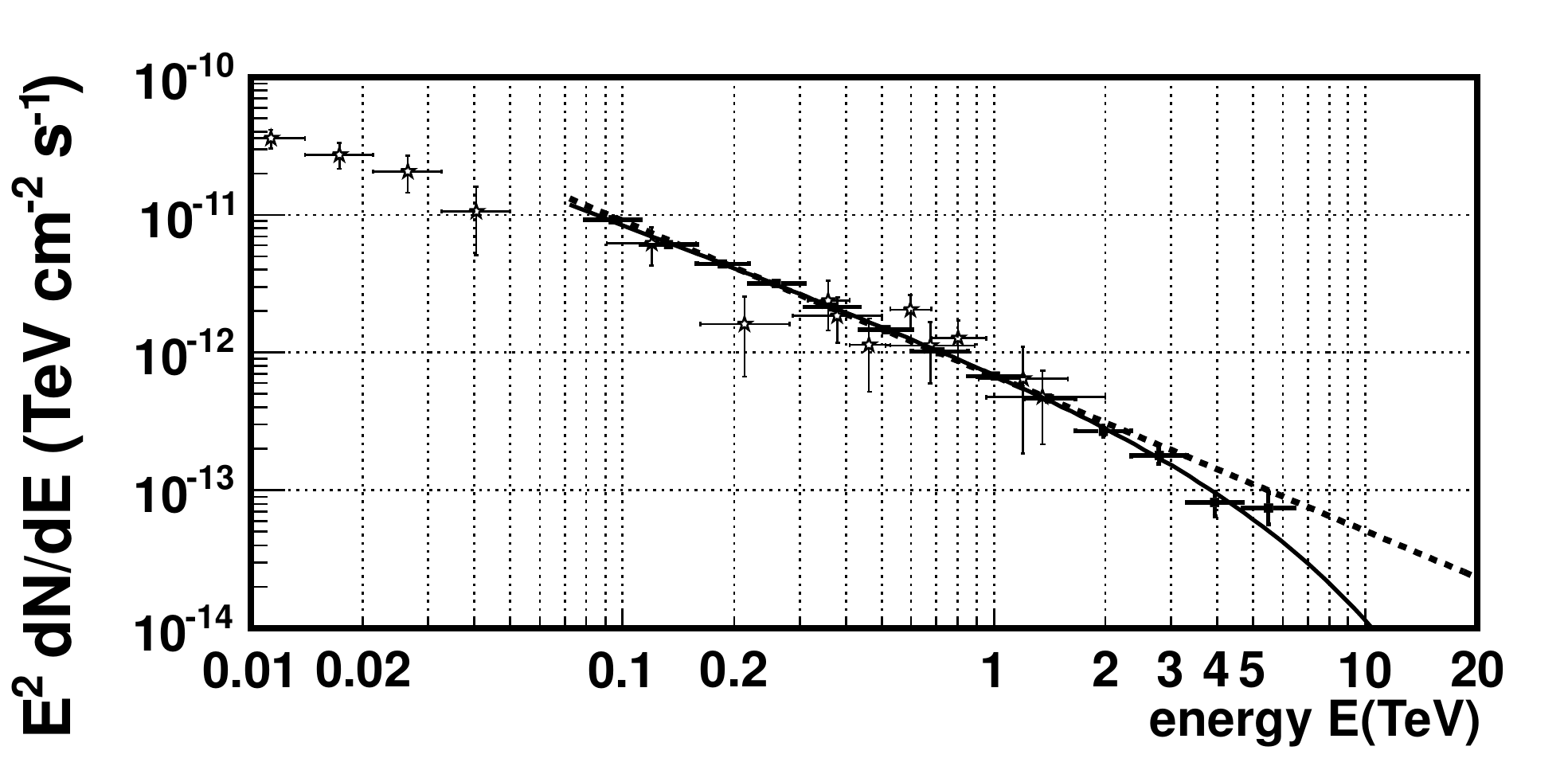}
\caption{The spectrum of IC443 is simulated (filled data points) with layout CTA-D for 50 hours of observation,
assuming an input spectrum in the form of a power law with an exponential cut off of 3 (top panel) and 5
(bottom panel) TeV. A power law index of $E^{-3}$ has been used. Fermi, VERITAS and MAGIC data points are also shown with open points.}
\label{IC443}
\end{figure}

However, if a source is less powerful (or with a steeper spectrum) than RX J1713.7-3946 it might not be
straightforward (even for CTA) to identify spectral features such as cutoffs/breaks if they
are located at very high energies, where the photon statistics are meager. In order to investigate this scenario, we consider here the SNR IC~443, which has been detected at both GeV \cite{IC443FERMI} and TeV \cite{IC443MAGIC,IC443VERITAS} energies and exhibits a quite steep spectrum.
It is important to stress here that for IC~443 the centroids of the GeV and TeV emission are misplaced, a fact that has been tentatively interpreted as the result of the escape of high energy CRs from the SNR shell and of their interaction with the ambient medium (for details see \cite{diegoIC443}). However, also in this scenario, the presence of a cutoff in the gamma-ray spectrum is expected in correspondence to the maximum energy of accelerated particles.

Here and in the following, in order to obtain the spectral points for a simulated object the following procedure is adopted. The energy domain of CTA is divided in logarithmically equally spaced energy bins. Excess and significances are calculated following \cite{liandma} (equation 5), with a fixed number of 5 background control regions ($\alpha$=0.2). For each bin, ON and OFF number counts are calculated from the intrinsic source spectrum convoluted with Monte Carlo simulated response functions of the array and from expected background rates (dominated by the cosmic ray background. These numbers are then randomized. A bin is kept when its significance passes the threshold of 3 standard deviations. Further cuts relate to the minimum level of excess above background in a bin (3\%) and the minimum number of excess counts (10).

In Figure~\ref{IC443} the
spectrum of IC~443 has been simulated for CTA in
configuration D and for 50 hours of exposure (filled data points). The input spectrum adopted is a power law with slope -3 with an exponential cutoff at 3 TeV (top panel) and 5 TeV (bottom panel). Also the data from MAGIC \cite{IC443MAGIC}, VERITAS \cite{IC443VERITAS}, and Fermi \cite{IC443FERMI} are indicated with open data points. The best power-law fit to MAGIC data provides a slightly different spectral slope equal to -3.1, consistent within errors with the spectrum measured by VERITAS, which is -2.99. The dotted lines represent a power-law fit to the simulated data, while the solid lines show a fit for a power law with an exponential cutoff. It can be seen from the figures that a cutoff can be identified in the spectrum only if it is at energies no greater than 5 TeV. 
For the top panel (cutoff at 3 TeV) the reduced-$\chi^2$ is 44.36/10 (fit probability $2 \times 10^{-6}$) and 2.87/9 (fit probability 0.96) for the power-law and power-law plus cutoff fits, respectively.
For the bottom panel (cutoff at 5 TeV) these quantities are 33.4/11 (fit probability $4 \times 10^{-4}$) and 4.9/10 (fit probability 0.89). 
In both cases the evidence for the cutoff is clear.
Conversely, for cutoffs located at higher energies the pure power-law model fits as well the data, and the position of the cutoff cannot be constrained. It is important to remark that with present TeV data it is not possible to perform such an accurate study on the spectral shape due to the larger error bars.

For brighter sources or for sources with a harder spectrum than IC~443 (e.g. RX J1713.7-3946), CTA will be able to determine with good accuracy the spectral shape of cutoffs/breaks or other spectral features (see e.g. \cite{matthieu}). Different CTA configurations will give a
different level of accuracy at low, intermediate, or high energies. We have chosen here
configuration D because it is the one which gives the best performances at high energies, i.e. it is the most optimistic scenario. To provide the reader with conservative numbers, we repeated the simulation by using configuration I, which constitutes a good compromise between the sensitivity at both low and high photon energies. In this case a discrimination between single power-law and power-law plus cutoff fits is still possible though quite less significant. If the cutoff is at 3 TeV the reduced-$\chi^2$ is 44.32/10 (fit probability $2.88 \times 10^{-6}$) and 0.287/9 (fit probability 0.96) for the power-law and power-law plus cutoff fits, respectively.
On the other hand, if the cutoff is at 5 TeV these quantities are 20.24/11 (fit probability $4.2 \times 10^{-2}$) and 12.13/10 (fit probability 0.28).

\subsection{Molecular clouds with CTA: probing the intensity of cosmic rays in the Galaxy and their diffusion properties}
\label{sec:SNRMC}

Massive Molecular Clouds (MC) are an important subject of investigation because they can provide a thick and massive target for CR hadronic interactions \cite{blackfazio,felixbarometers}. For this reason, MCs are expected to emit gamma rays and, indeed, the emission observed from some MCs in the Galaxy has been interpreted as the result of the decay of neutral pions produced by CR protons interacting with the dense gas that constitutes the cloud (e.g. the MCs in the galactic centre region \cite{HESSridge}, the ones close to SNRs such as W28 \cite{W28} or IC~443 \cite{IC443MAGIC}, just to mention the most prominent cases). The gamma-ray luminosity of the MC would be proportional to the cloud mass and to the CR intensity in the cloud.

For these reasons, the search for gamma-ray emission from MCs with measured mass and distance has been proposed as a tool to probe the CR intensity at the cloud's location (e.g. \cite{felixbarometers,issa,sabrinabarometers}).
If the MC is embedded in the galactic CR background (sometimes referred to as {\it CR sea}), then the expected gamma-ray flux is proportional to the quantity $M_{cl}/d^2$, where $M_{cl}$ is the cloud mass and $d$ is its distance, and the MC is called {\it passive}.
Since the CR background is known not to vary a lot throughout the Galaxy \footnote{ Theoretical expectations for the spatial fluctuations of CR density in the Galaxy are of the order of $\approx 10$\% for TeV CRs \cite{ptuskinfluctuations}. Thus our assumption of a constant background of CR is very well justified.}, the detection of a MC in gamma rays with a flux significantly larger than the one expected for a passive cloud of the same mass would indicate that a source of CRs is present in the vicinity of or inside the cloud. Thus, gamma-ray-bright MCs can be used, in principle, to locate the sources of CRs \cite{montmerle,morfill,adv,atoyan,rodriguezmarrero,gabici09}. 

For these reasons, an investigation of the capability of CTA to detect MCs is of paramount importance. To this purpose, we consider four different situations: {\it i)} a passive MC (i.e. with no CR accelerator in its proximity); {\it ii)} a CR accelerator inside a MC; {\it iii)} a MC illuminated by CRs from a nearby accelerator; {\it iv)} a CR accelerator located in a region filled with dense, diffuse gas.

\subsubsection{Passive molecular clouds}

\begin{figure}
\centering
\includegraphics[width=.7\textwidth]{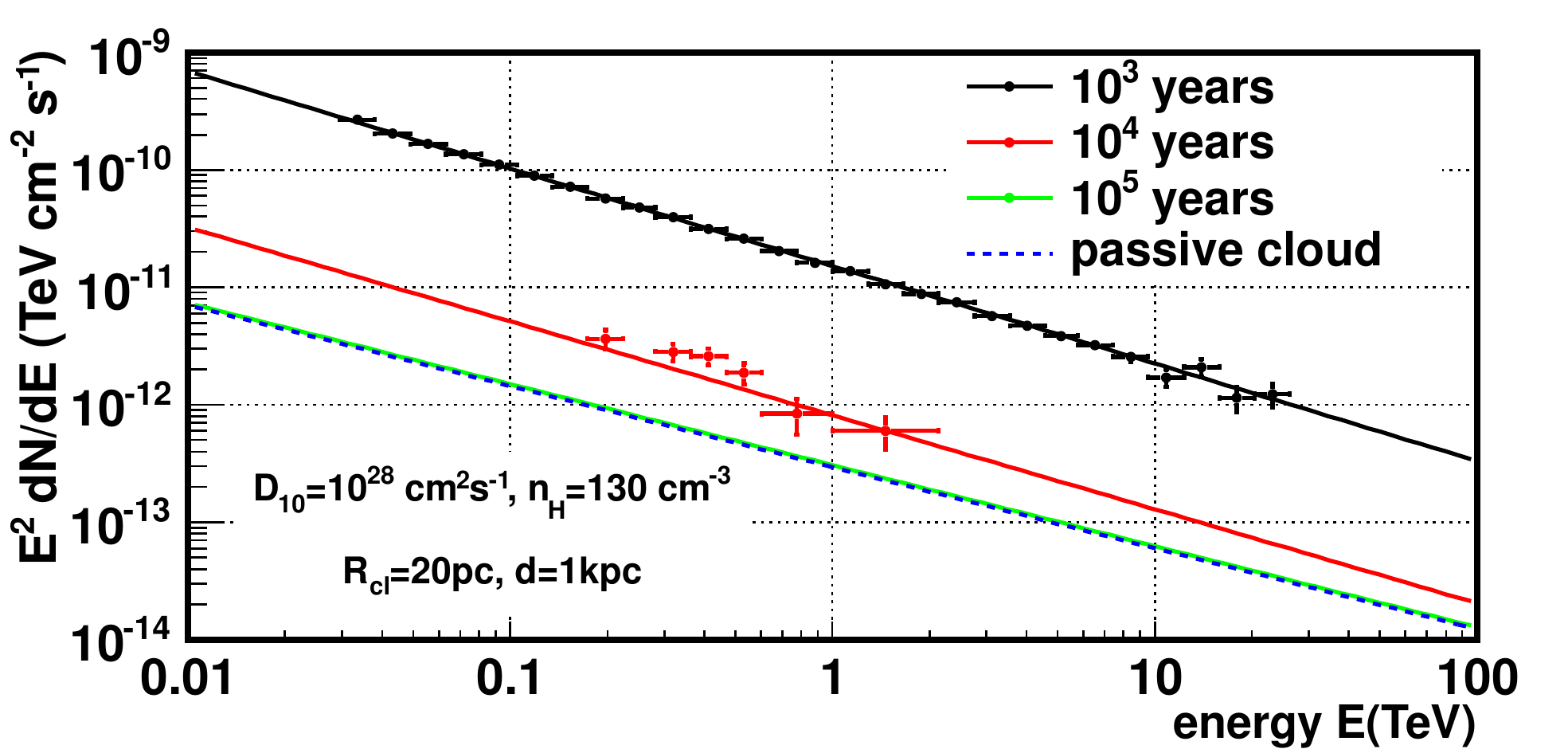}
\includegraphics[width=.7\textwidth]{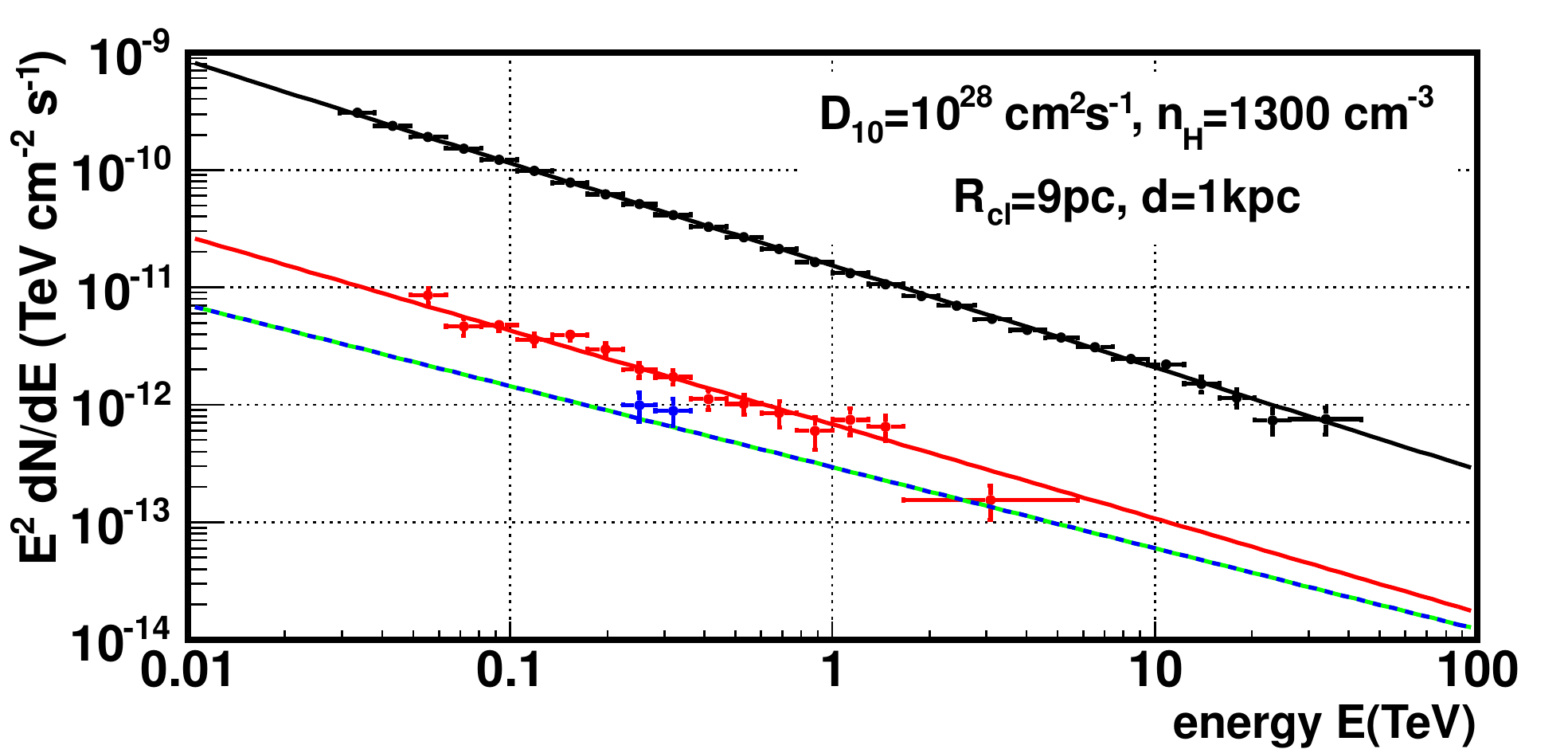}
\includegraphics[width=.7\textwidth]{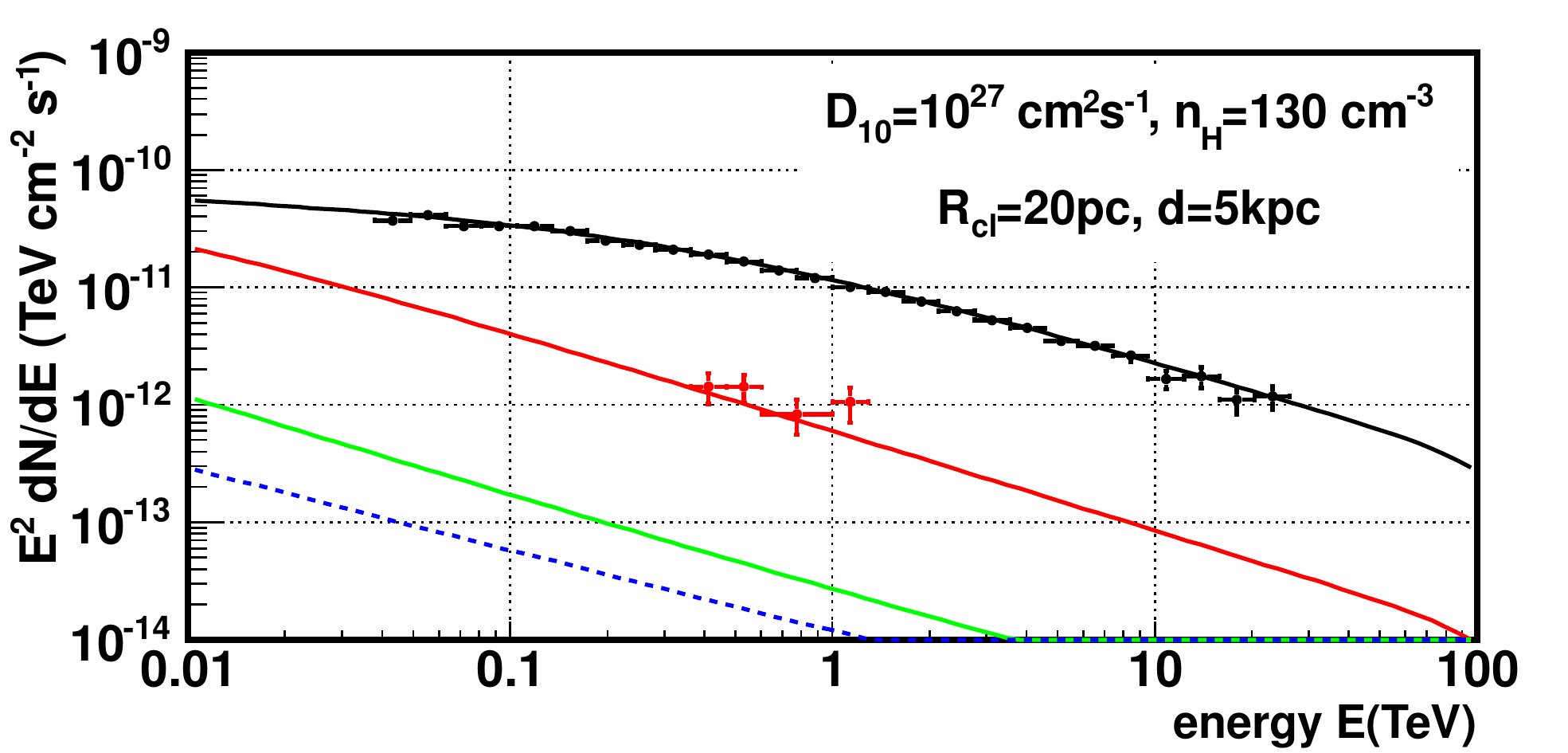}
\caption{Gamma-ray emission from a molecular cloud of mass $10^5 M_{\odot}$ with a CR accelerator located at the cloud's centre. The accelerator releases instantaneously $10^{50}$erg of CRs with a spectrum $\propto E^{-2.2}$. Black, red, and green lines represent the gamma-ray flux for $10^3$, $10^4$, and $10^5$ years after the release of CRs. The dotted blue line represents the emission from a passive cloud. Data points are simulated CTA observations for 50 hours of exposure. TOP PANEL: cloud density and radius are 130 cm$^{-3}$ and $\sim 20$ pc, respectively. Diffusion coefficient is $D = 10^{28} (E/10~{\rm GeV})^{0.5}$ cm$^2$/s. MIDDLE PANEL: same as in the top panel, but cloud density and radius are 1300 cm$^{-3}$ and $\sim 9$ pc. BOTTOM PANEL: same as in the top panel except for the cloud distance (5 kpc) and for the diffusion coefficient which is reduced by a factor of 10.}
\label{giovanna}
\end{figure}

Consider a cloud with mass $M_5 = M/10^5 M_{\odot}$ located at a distance $d_{kpc} = d/{\rm kpc}$. If the cloud is embedded in the CR background, with an intensity of $J_{bg} \approx 2.2 ~ E^{-2.75} {\rm cm^{-2} s^{-1} sr^{-1} GeV^{-1}}$ its gamma-ray flux due to proton-proton interactions Å can be calculated and has an energy spectrum as shown in Figure~\ref{giovanna} with a dotted blue line (for $M_5 = d_{kpc} = 1$). All the other curves in Figure~\ref{giovanna} will be described in the next Section. The flux of the passive MC at 1 TeV is roughly $\approx 3 \times 10^{-13}~{\rm TeV/cm^2/s}$, roughly 3 times above the CTA sensitivity for a point-like source and 50h of observation. However, the angular extension of MCs is expected to be much larger than the angular resolution of Cherenkov telescopes. Thus a more detailed discussion is needed in order to assess the capability of CTA to detect such extended objects. 

Let us then estimate the expected angular extension of a MC and compare its flux with the instrumental sensitivity for a source with that extension. If $\bar{n}_3$ is the average cloud density in units of 1000 protons per cubic centimeter, then the angular size (radius) of the cloud is $\vartheta_{cl} \approx 0.6^{\circ} d_{kpc}^{-1}  (M_5/\bar{n}_3)^{1/3}$. For a passive cloud with homogeneous density, the 68\% containment radius of the gamma-ray emission is given by $\vartheta_{68\%} \approx 0.7 ~ \vartheta_{cl}$, which for the values considered here is approximately 0.4$^{\circ}$. This has to be compared with the angular resolution of CTA that is expected to be of the order of $\vartheta_{CTA} \approx 0.05^{\circ}$ \cite[see e.g.][]{matthieu}. The sensitivity of a Cherenkov telescope for an extended source scales approximately as the ratio between the source size and the instrumental angular resolution, and is thus a factor of $\vartheta_{68\%}/\vartheta_{CTA} \approx 8$ worse than the sensitivity for a point source, resulting in a sensitivity (very roughly) of the order of $\approx 10^{-12}$~TeV/cm$^2$/s. Thus, from these considerations and from Figure~\ref{giovanna} one can see that the detection of a passive cloud is challenging even for an array as sensitive as CTA.
Results from a detailed study are reported in Figure~\ref{giovanna} and demonstrate that a dense ($\approx 10^3$cm$^{-3}$) cloud of mass $10^5 M_{\odot}$ will be marginally detectable at sub-TeV energies if located at a distance of $\approx 1$~kpc (Figure~\ref{giovanna}, middle panel).
A stacking approach (i.e. summing the pointed observations of several nearby MCs) might improve the chances of a detection. 

A number of MCs with masses in the range $10^4...10^5 M_{\odot}$ are located well within a kiloparsec from the Earth \cite{dame1,dame2}. As an example, detailed studies of massive clouds in the Perseus, Taurus, and Auriga region can be found in \cite{ungerechts}, while for the Orion-Monoceros complex we refer to \cite{maddalena, wilson}. Even though these MCs have masses in the relevant range, they are very often very extended (several degrees) making their individual detection at TeV energies problematic. 

\subsubsection{Cosmic ray accelerator inside a molecular cloud}

If a CR accelerator is located inside a MC, the expected gamma-ray flux from the cloud is enhanced due to the presence of the accelerated CRs. In Figure~\ref{giovanna} the situation in which an accelerator is releasing instantaneously, i.e. over a time scale much shorter than the CR diffusion time, $10^{50}$~erg of CRs is considered. The spectrum of these runaway cosmic rays is assumed to be a power law of the form $\propto E^{-2.2}$. In the top panel, CRs are assumed to diffuse away from their accelerator with a (homogeneous and isotropic) diffusion coefficient given by $D = 10^{28} (E/10~ {\rm GeV})^{0.5}~{\rm cm^2/s}$. This assumption for $D$ is consistent with CR data (e.g. \cite{CRbook}).
The assumed MC gas proton density is 130 cm$^{-3}$, implying, for a total gas mass of $10^5 M_{\odot}$ and for a homogeneous gas distribution, a radius of $\sim 20$~pc for the cloud. The CR accelerator is located at the centre of the MC.
The black, red, and green lines represent the gamma-ray spectrum from the cloud after $10^3$, $10^4$, and $10^5$~yr, respectively. Spectra have been calculated following \cite{atoyan}.
Note that for times of the order of $\sim 10^5$~yr the MC emission coincides with that expected from a passive cloud, represented by the dotted blue line. This indicates that CRs that escaped the accelerator are diluted over a large volume, and thus their contribution to the cloud emission is negligible if compared with the contribution from the CR background.
Data points represent simulated observations with CTA, configuration E, for 50 hours of exposure.
For this choice of parameters, the cloud is clearly detectable for $\sim 10^4$~yr after the release of CRs from the accelerator.
By comparing the red and the blue (i.e. a passive cloud) lines, we can conclude that an overdensity of CRs above the galactic sea of a factor of a few suffices to make a nearby ($d \sim 1$~kpc) MC of $10^5$ solar masses detectable by CTA. Larger CR overdensities and/or a larger cloud mass are necessary to detect a cloud at distances significantly beyond 1 kpc.
Such large overdensities can be obtained for shorter times, of the order of $\sim 10^3$~yr (e.g. \cite{atoyan}).

The middle panel of Figure~\ref{giovanna} is identical to the top panel, except for the fact that the cloud density is enhanced by a factor of 10, to get $n_{cl} \sim 1.3 \times 10^3~{\rm cm^{-3}}$. This corresponds to a cloud radius of $\sim 9$~pc. Thus, the angular extent of the gamma-ray emission is smaller, making a detection easier.
This is indicated by the smaller error bars in the simulated data, and by the fact that the cloud is now detected over a broader energy range, if compared to the cloud considered in the top panel.

Finally, in the bottom panel of Figure~\ref{giovanna}, all the parameters are set as in the top panel (i.e. gas density $\sim 130$~cm$^{-3}$, cloud mass $10^5 M_{\odot}$) except for the diffusion coefficient which is now one order of magnitude smaller: $D = 10^{27} (E/10~ {\rm GeV})^{0.5}~{\rm cm^2/s}$ and for the fact that the cloud is now assumed to be at a distance of 5 kpc. A diffusion coefficient significantly smaller than the average galactic one might be justified close to CR accelerators since runaway CRs can enhance the level of magnetic turbulence due to streaming instability and in turn suppress the diffusion coefficient (see e.g. \cite{plesser}).
If the diffusion coefficient is small, CRs are confined more effectively within the MC.
As a consequence, the resulting gamma-ray fluxes are significantly enhanced, and the cloud is detectable up to significantly larger distances. For the case considered here, $d = 5$~kpc, the cloud is detectable for an age of $\lesssim 10^4$~yr after the release of CRs.
Moreover, for this choice of the parameters, a break appears in the spectrum, and its position corresponds to the CR particle energy that satisfies the relation: $R_{cl} \sim \sqrt{6~D(E)~t}$.
This is the energy of the particles that in a given time $t$ are able to diffuse up to the cloud's border.
Thus, at time $t$, all the particles with lower energy are confined inside the cloud and thus the corresponding gamma-ray spectrum simply mimics the CR injection spectrum, $\propto E^{-2.2}$.
On the other hand, above that energy CRs begin to leak away from the MC. The escape is more pronounced at higher energies and thus the resulting spectrum is steeper than the injection one.
The break energy moves to lower and lower energies with time (in the case considered here as $\propto t^{-2}$) as CRs with lower and lower energy leak out of the cloud.
This is why the break is only barely visible in the red curve ($t = 10^4$~yr) and absent in the green one ($t = 10^5$~yr).
The effective confinement of CRs inside the cloud enhances the gamma-ray emission and makes the molecular cloud detectable at much larger distances than the ones considered in the top and middle panel.
A detailed treatment and an extended discussion of the scenario presented in this section will be discussed elsewhere \cite{giovanna}.


\subsubsection{Gamma-ray emission from a molecular cloud illuminated by cosmic rays from a nearby accelerator}

Another scenario that needs to be investigated is the one in which a MC is located at a certain distance from a CR accelerator. CRs escaping from the accelerator can then illuminate the MC and produce gamma rays due to proton-proton interactions. Notably, besides being an indirect way to identify CR sources, studies of illuminated MCs may serve to estimate the CR diffusion coefficient in the vicinity of CR accelerators, since the properties of the expected gamma-ray emission depend on the exact value and on the energy dependence of the diffusion coefficient.
For a general discussion of this scenario see \cite{atoyan}. In a more recent study, where SNRs were considered as the illuminating sources, it was realized that the superposition of the gamma-ray emission from background CRs (with a steep spectrum) and CRs coming from the nearby CR source (with a hard spectrum) could produce characteristic concave, or V-shaped gamma-ray spectra \cite{gabici09}. The detection of such V-shaped spectra would be a very important achievement since it would allow to test the generally accepted ideas about CR escape from sources and their propagation away from the production site.

\begin{figure}
\centering
\includegraphics[width=.7\textwidth]{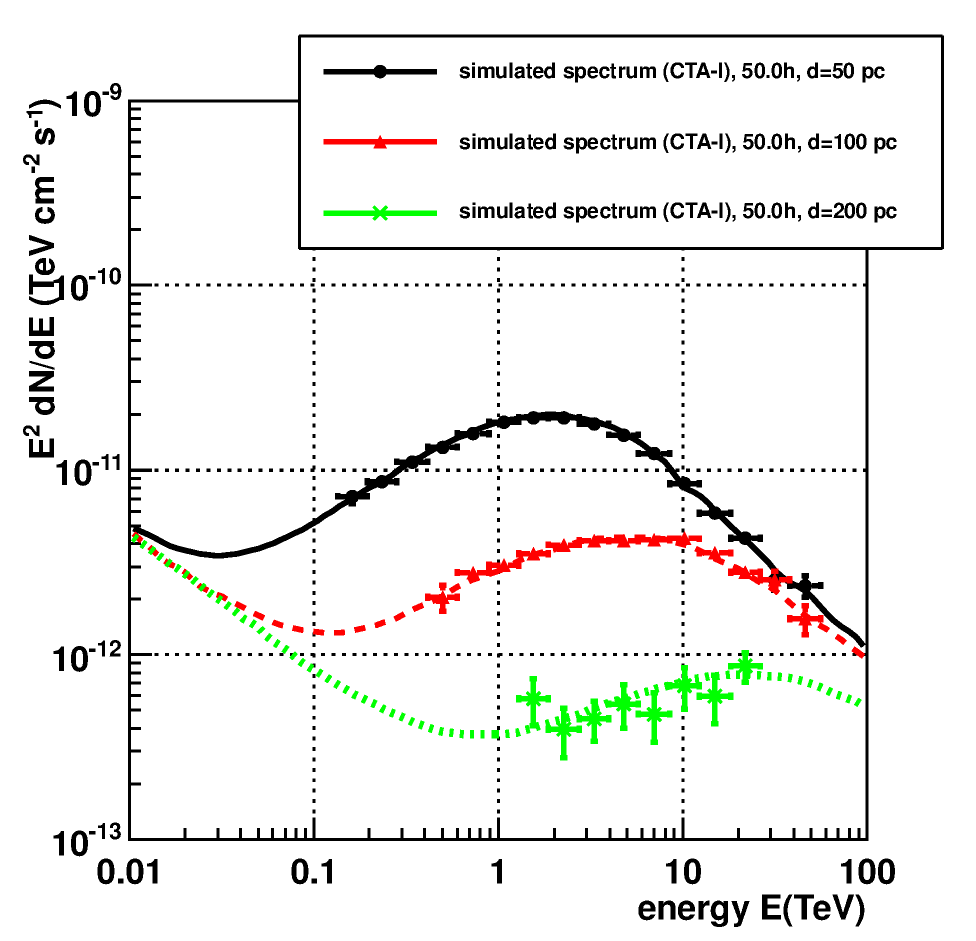}
\caption{Gamma-ray spectra for a MC illuminated by CRs coming from a nearby SNR (lines) and simulated observations (data points) for 50 hours for CTA configuration I. Black solid, red dashed and green dotted lines refer to distances between the SNR and the MC of 50, 100, and 200 pc, respectively. The distance of the MC is 1 kpc, cloud mass is $10^5 M_{\odot}$ and SNR age is 2000 years. }
\label{elsailluminated}
\end{figure}

In order to test the feasibility of a detection of such sources, we plot in Figure~\ref{elsailluminated} the expected gamma-ray emission from a cloud of $10^5 M_{\odot}$ at a distance of 1 kpc. A SNR emitting $3 \times 10^{50}$ erg in form of CR is located at 50 (black solid), 100 (red dashed), and 200 pc (green dotted line) from the MC. Data points are simulated CTA observations of 50 hours in configuration I. The age of the SNR is 2000 years and a CR diffusion coefficient equal to $10^{28} (E/{\rm GeV})^{0.5}$ has been used in the calculations. Details of the modeling can be found in \cite{gabici09}. The expected spectra exhibit a concave shape, with minimum energy $E_*$ (in $E^2 F(E)$) that varies in a broad energy range between tens of GeVs to TeVs. In Figure~\ref{elsailluminated} the difference in the energy of the concavity is due to the varying distances to the SNR, but also a difference in the SNR age, or a different assumption for the value of the diffusion coefficient would produce a shift in $E_*$ (for details see \cite{gabici09}). To date, no such spectra have been observed, despite the relatively high fluxes expected. One possible explanation for this is the fact that such emission, exhibiting a clear V-shaped spectrum, is expected to last $\lesssim 10^4$ years only (under standard assumptions for the value of the diffusion coefficient). After that $E_*$ moves into the GeV range and the total spectrum resembles a power law. Thus, MCs with a V-shaped spectrum might be rare, distant objects, with fluxes below the sensitivities of currently operating instruments.

From Figure~\ref{elsailluminated}  we can see that illuminated massive clouds could be easily detected if located at 1 kpc from the Earth. A MC which is quite close to the SNR (e.g. 50 pc, as for the black solid line in Figure~\ref{elsailluminated}) could be detected up to roughly the distance of the Galactic centre, thing that would be hardly possible with current telescopes. However, despite the high fluxes expected, detecting the concavity in the spectrum will be difficult even for an instrument as sensitive as CTA. The reason is that the dip of the concavity is in most cases located at low energies, close to or below the smallest ones probed by CTA. This suggests that studies that will combine observations with CTA with Fermi observations in the GeV range will be the most powerful to investigate the scenario of illuminated MCs.

\subsubsection{Gamma-ray emission from interactions of runaway cosmic rays in a diffuse gas surrounding the accelerator}

Finally, we consider the case of a CR accelerator located in a relatively dense and diffuse (i.e. not a massive MC) interstellar medium. Once CRs are released from the acceleration site, they diffuse away in the surrounding medium and produce a diffuse emission of gamma rays due to proton-proton interactions. The detection of such emission might serve to identify the source of CRs and, under certain assumptions (i.e. on the source age and distance), to constrain the properties of CR diffusion in the vicinity of the accelerator. To investigate this scenario,  we take as an example the SNR RX J1713-3946, for which a detailed study on CR escape and subsequent diffusion and related gamma-ray emission exist \cite{sabrinaRXJ}.

\begin{figure}
\centering
\includegraphics[width=.8\textwidth]{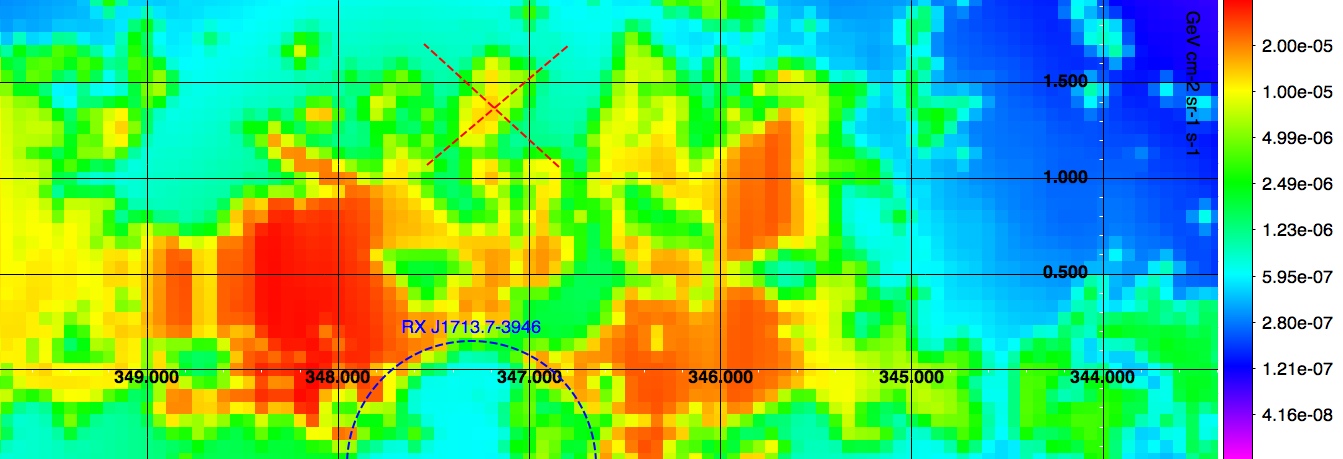} 
\includegraphics[width=0.99\textwidth]{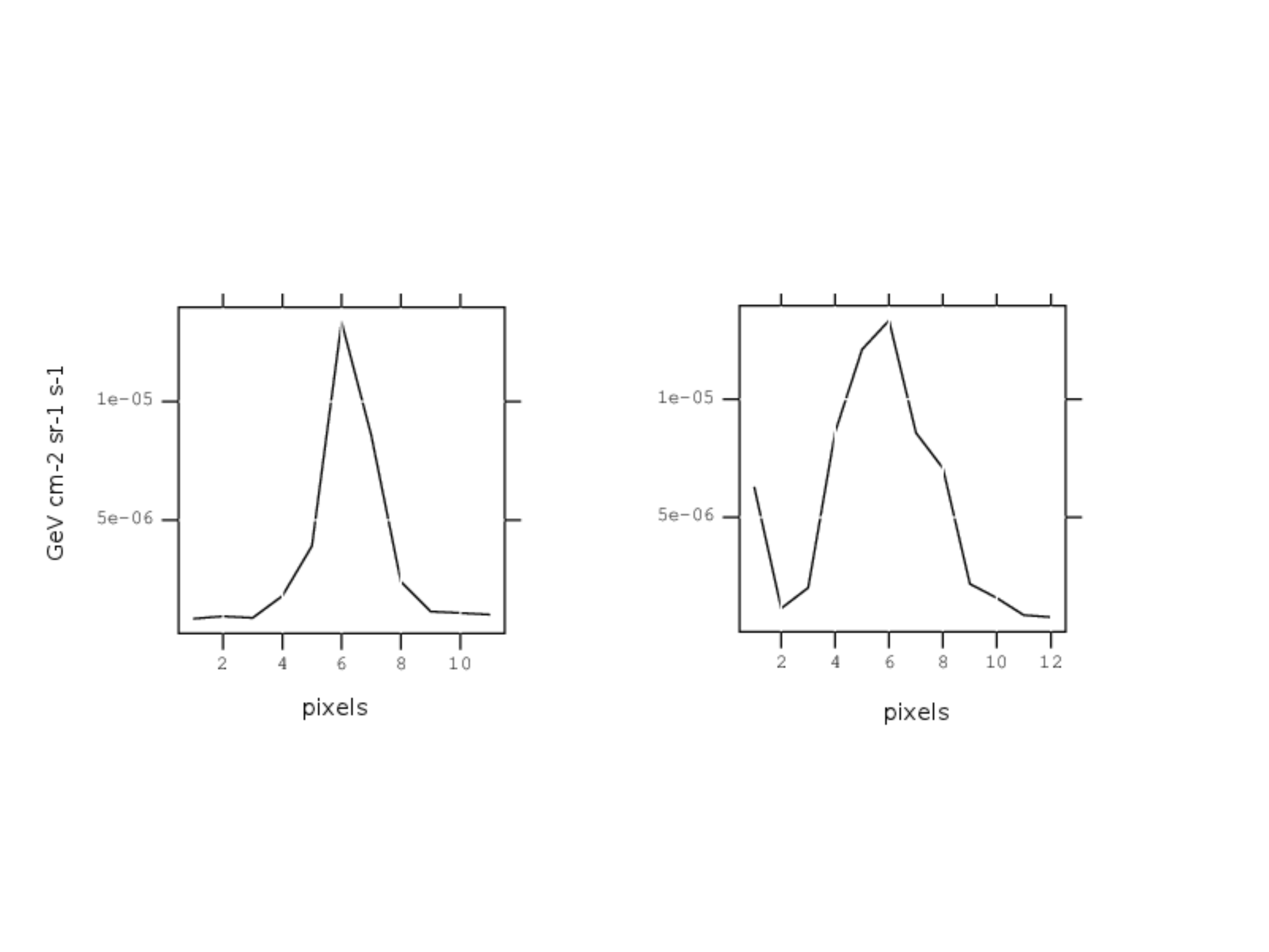} 
\caption{Top panel: diffuse gamma-ray emission from the region surrounding the SNR RX J1713-3946 as calculated in \cite{sabrinaRXJ}. The position of the SNR shell is indicated in purple. Units in the color scale are GeV/cm$^2$/s/sr.
Bottom panels: brightness profile (same units) of the emission along the two black lines in the map. The x-axis is in pixel units. One pixel is 0.066 degrees.}
\label{sabrina}
\end{figure}

The assumptions made in \cite{sabrinaRXJ} are the following: CR protons are released gradually in time from the SNR shock, the one with the highest energy (assumed to be $\approx 500$~TeV) at the end of the free expansion phase of the SNR evolution (i.e. $\approx 100$~yr), and those with lower energies later in time according to the relationship $E_{esc} \approx 500 (t/100~{\rm yr})^{-\delta}$~TeV, with $\delta = 0.43$. The CR acceleration efficiency at the SNR shock is assumed to be 30\% of the total supernova explosion energy, for which the standard value $10^{51}$ erg has been considered. This is the maximum amount of (hadronic) CRs that can be considered without violating the GeV observations that suggest that the observed gamma-ray emission from RX J1713-3946 is probably of leptonic origin \cite{fermiRXJ}. The spatial distribution of CRs around the source is calculated by solving the diffusion equation with an isotropic diffusion coefficient of the form: $D = D_0 (E/10 ~ {\rm GeV})^{0.5}$. The diffuse gamma-ray emission is calculated following \cite{kelner} and using the gas density distribution inferred from observations of the CO molecular line (obtained with the NANTEN telescope, \cite{nanten} and references therein) and of the H${\sc I}$ line (as in the LAB survey, \cite{kalberla}). In \cite{sabrinaRXJ}, the distance of the system (SNR+gas) was assumed to be 1 kpc. A map of the resulting diffuse emission for $D_0 = 10^{28}$ cm$^2$/s is shown in Fig.~\ref{sabrina}.

\begin{figure}
\centering
\includegraphics[width=.8\textwidth]{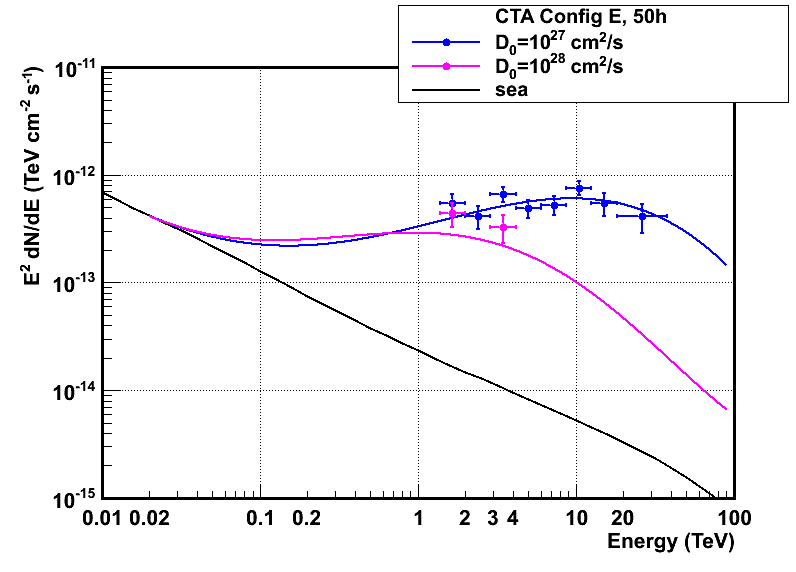}
\caption{Gamma-ray flux from the gas clump indicated by a black cross in the map shown in Fig.~\ref{sabrina}. The curves represent the predicted emission and the data points are the simulated CTA observations (configuration E) for 50 hours of exposure.}
\label{francesca}
\end{figure}

In Fig.~\ref{francesca} the gamma-ray spectrum is plotted from the small gas clump indicated in the map by the black cross. The blue and magenta curves refer to $D_0 = 10^{27}$ and $10^{28}$ cm$^2$/s, respectively. The brightness distributions along the two black lines is plotted in the two insets in Fig.~\ref{sabrina}.  Thus, the data points in Fig.~\ref{francesca} have been obtained by approximating the emission from the clump with a Gaussian distribution with $\sigma \sim 0.15^{\circ}$ and for 50 hours of observation for CTA configuration E. As can be seen, the flux levels are within the capabilities of CTA, though the hard gamma-ray spectra expected in this scenario might be detected at significantly only at high energies (i.e. $\gtrsim 1$ TeV).

It must be noted that the study of the diffuse emission is a very difficult task for Cherenkov telescopes. An accurate background subtraction has to be performed, and a strategy for mapping regions of size comparable with the instrument field of view needs to be implemented. In producing Fig.~\ref{francesca} we didn't take into account the bright emission from RXJ1713.7-3946, nor we performed a complete background subtraction that takes into account the diffuse excess of gamma rays in the region surrounding the simulated clump.
For these reasons, the results reported here have to be considered as qualitative, but still encouraging, hints for the fact that CTA will be able to perform studies of the diffuse emission surrounding CR sources and possibly will be capable of resolving the details of such emission. Besides locating the sites of CR acceleration in the Galaxy, these studies will also serve to investigate the transport of CRs at specific locations in our Galaxy, an aspect that is still at the limit of the capabilities of current instruments.

\section{Starburst galaxies and galaxy cluster}

Although proposed and predicted as high-energy gamma-ray emitters over two decades ago (see e.g. \cite{paglione}), starburst galaxies were finally established through VHE- \cite{HESSstarburst,VERITASstarburst} and GeV observations \cite{FERMIstarbursts} only recently. Although the two archetypal starburst galaxies, M82 and NGC253, appear as point sources to current instruments, it became clear that the starburst phenomenon relates to the core of the systems, where a potentially high CR density is sustained by higher star formation rates and greater amount of gas and dust that reprocess light into the IR band and can then serve as targets for gamma-ray production by CR electrons and nuclei. Due to the large density of targets (both gas and photons) for CR interactions, which diminishes the CR energy loss time, the conversion efficiency of protons into gamma-rays exceeds that in our Milkyway by up to an order to magnitude. To date, the scientific return beyond the physics scenarios that led to the prediction of their high-energy emission is still limited, as the spectrum is only coarsely measured and, accordingly, no observations exist that could address the question of the maximum energy, or the potential existence and shape of spectral cutoffs. A detailed spectral study will be in reach with CTA.

\begin{figure}
\centering
\includegraphics[width=.8\textwidth]{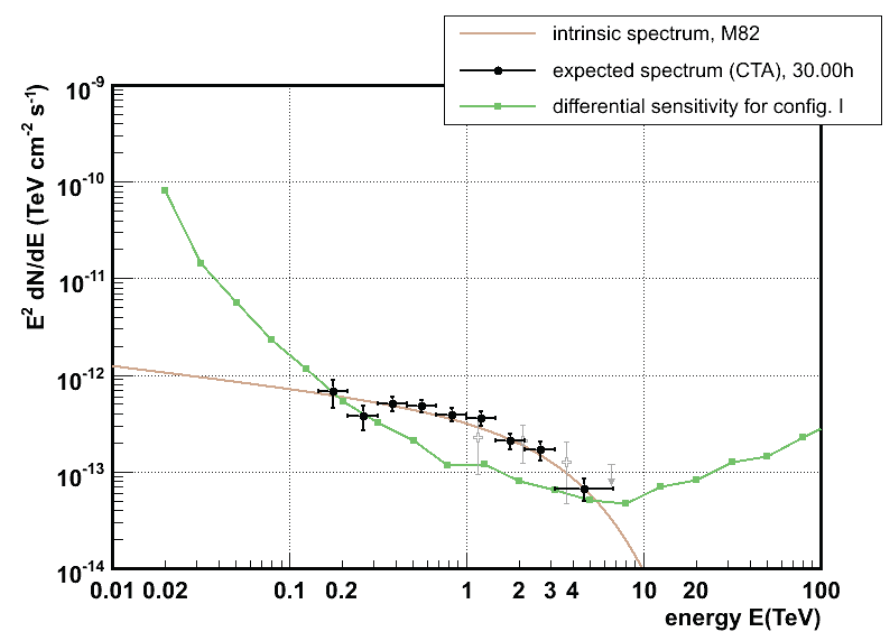}
\caption{Simulation of 30h CTA observations (array configuration I) for the starburst galaxy M82. The curves represents the predicted emission and the data points based on the measured spectrum from M82 and a spectral model of \citep{M82model}.}
\label{elsa}
\end{figure}

This is demonstrated in Figure~\ref{elsa}, where the simulated spectrum for M82 is shown (black data points) for 30 hours of observations with CTA in configuration I. Grey data points refer to VERITAS observations. The simulated spectrum is based on the model presented in \cite{M82model}. It is evident that, if present, a cutoff in the TeV energy range will be easily identified by CTA.

In contrast to starburst galaxies, the detection of high energy gamma-ray emission from galaxy clusters still eludes present-day GeV (AGILE, Fermi) and TeV (H.E.S.S., MAGIC, VERITAS) observatories and experiments. There are, however, good reasons to assume that clusters of galaxies emit high-energy gamma-radiation: Galaxy cluster are evolving massive structures in the Universe that should dissipate certain amounts of energy through (merger and accretion) shocks, winds or turbulence, and they host objects like Active Galactic Nuclei and/or radio galaxies, which are both prone for relativistic outflows and accordingly relevant for particle injection and energetic feedback in galaxy clusters. 
With the anticipated sensitivity of CTA, the minimal hadronic model for the gamma-ray flux in a galaxy cluster \cite{PerseusMagic} will be testable. So far, the deep exposure obtained by the MAGIC telescope of the Perseus cluster \cite{PerseusMagic} constitute the most stringent constraints from gamma-ray observations regarding CR energy ($E_\mathrm{CR} /E_\mathrm{th} < 5...3\%$), CR-to-thermal pressure
($\langle X_\mathrm{CR} \rangle < 8\%$ for the entire cluster and $ < 4\%$ for the core), maximum shock acceleration efficiency, and value of the magnetic field in a galaxy cluster. Since the upper limits are only a factor of $\sim 2$ larger than the model prediction for the CR-induced $\gamma$-ray emission \cite{PerseusMagic} , the discovery potential to finally detect a galaxy cluster at gamma-rays is clearly given, but also a non-detection at CTA-sensitivity level would imply CR fluxes that are way too small to produce sufficient electrons through hadronic interactions with the ambient matter to explain the observed synchrotron emission - thus bringing yet another very intriguing science aspect into cluster research.

\section{Conclusions}

In this paper we discussed the expected impact of future observations with the Cherenkov Telescope Array for cosmic ray studies. 
With its improved sensitivity, CTA is expected to significantly increase the number of detected sources. This will make population studies of SNRs possible, which are to date the most promising candidate sources for CR acceleration in the Galaxy. Being able to detect such objects up to the other side of the Galaxy, CTA is expected to provide us with a sample of several tens of TeV-bright young SNRs (see Section 2.1). 
Also, its improved sensitivity over a large energy range will substantially increase the quality of spectral studies. This will possibly allow the detection of cutoffs or breaks in gamma-ray spectra of SNRs, which to date has been possible for the bright and best studied SNR RX~J1713.7-3946 only. These studies will shed light on the details of the mechanism of particle acceleration at shocks, especially for what concerns the maximum energy of the accelerated particles. 

Also their propagation of CRs in the interstellar turbulent magnetic field needs to be understood. Gamma-ray observations of the regions surrounding SNRs (or any other source of CRs) are an important tool to constrain the diffusive behavior of CRs. This is because CRs after leaving their sources produce gamma rays in interactions with the ambient gas surrounding the accelerator. With its large field of view and improved sensitivity, CTA is expected to be able to detect this diffuse emission surrounding nearby ($\sim 1$ kpc) and powerful (a few $10^{50}$ erg) sources of CRs. If massive $\sim 10^5 M_{\odot}$ MCs are located in the vicinity of the CR source, the expected emission is at a level detectable even for large distances of the MC, comparable with the distance of the galactic centre. This seems to ensure that CTA will significantly increase the number of MCs detected at TeV energies.

After escaping the sources, CRs diffuse in the turbulent magnetic field of the Galaxy and finally escape into the intergalactic medium. For this reason, both galaxies and clusters of galaxies are expected to emit gamma rays due to CR interactions with the interstellar and intracluster gas and radiation field. Starburst galaxies, with an enhanced star formation rate and high density and radiation field, are ideal target for observations. Though to date only two starburst galaxies and no clusters of galaxies have been detected in the TeV domain, the chances for more detections seems favorable. In the case of clusters of galaxies, a detection would establish a new class of TeV sources. Since starburst galaxies and clusters of galaxies are the storage rooms of CRs after their acceleration, detecting and understanding their gamma-ray emission will indirectly tell us something about the nature and efficiency of CR sources and will also help us in constraining the characteristics of CR propagation in turbulent magnetic fields.


\section*{Acknowledgements}

We gratefully acknowledge support from the agencies and organizations listed in this page: http://www.cta-observatory.org/?q=node/22 .
We thank J. Conrad, E. de Gouveia Dal Pino, S. Digel, W. Hofmann, D. Horns, B. Khelifi, and R. Yamazaki who provided comments on the manuscript.
S.G. acknowledges support from the EU [FP7 - grant agr. n$^o$256464].
E. dC., D.H., G.P. and D.F.T. acknowledge support from the Ministry of Science and the
Generalitat de Catalunya, through the grants AYA2009-07391 and
SGR2009-811, as well as by ASPERA-EU through grant EUI-2009-04072.


\bibliographystyle{model1a-num-names}







\end{document}